\documentclass[11pt,a4paper]{article}
\usepackage{jheppub}
\usepackage{subcaption}

\usepackage{graphicx}
\usepackage{dcolumn}
\usepackage{bm}
\usepackage{amsmath}
\usepackage{mathrsfs}
\usepackage{multirow}
\usepackage{xcolor}

\newcommand{\gsim}{\gtrsim}
\newcommand{\lsim}{\lesssim}

\newcommand{\rd}{\partial}
\newcommand{\nn}{\nonumber}

\newcommand{\mca}{\mathcal{A}}
\newcommand{\mcb}{\mathcal{B}}
\newcommand{\mcc}{\mathcal{C}}
\newcommand{\mcd}{\mathcal{D}}
\newcommand{\mcf}{\mathcal{F}}

\newcommand{\mco}{\mathcal{O}}
\newcommand{\mcq}{\mathcal{Q}}
\newcommand{\mct}{\mathcal{T}}
\newcommand{\mcu}{\mathcal{U}}

\renewcommand{\lg}{\mathscr{L}}

\newcommand{\br}{{\rm B}}

\newcommand{\sm}{{\rm SM}}

\newcommand{\fb}{{\,{\rm fb}}}

\newcommand{\gev}{{\;{\rm GeV}}}
\newcommand{\tev}{{\;{\rm TeV}}}

\newcommand{\beq}{\begin{equation}}
\newcommand{\eeq}{\end{equation}}
\newcommand{\bea}{\begin{eqnarray}}
\newcommand{\eea}{\end{eqnarray}}
\newcommand{\barr}{\begin{array}}
\newcommand{\earr}{\end{array}}
\newcommand{\bc}{\begin{center}}
\newcommand{\ec}{\end{center}}
\newcommand{\bit}{\begin{itemize}}
\newcommand{\eit}{\end{itemize}}
\newcommand{\ben}{\begin{enumerate}}
\newcommand{\een}{\end{enumerate}}

\newcommand{\al}{\alpha}
\newcommand{\bt}{\beta}

\newcommand{\dt}{\delta}
\newcommand{\Dt}{\Delta}

\newcommand{\sg}{\sigma}

\newcommand{\kp}{\kappa}

\newcommand{\gm}{\gamma}
\newcommand{\Gm}{\Gamma}
\newcommand{\lm}{\lambda}
\newcommand{\mn}{{\mu\nu}}

\newcommand{\sthf}      {s_{\theta_F}}
\newcommand{\cthf}      {c_{\theta_F}}
\newcommand{\stthf}      {s_{2\theta_F}}

\newcommand{\sthu}      {s_{\theta_\mathcal{U}}}
\newcommand{\cthu}      {c_{\theta_\mathcal{U}}}

\newcommand{\sthd}      {s_{\theta_\mathcal{D}}}
\newcommand{\cthd}      {c_{\theta_\mathcal{D}}}

\newcommand{\gh}{\hat{g}}

\newcommand{\rr}      {{\gamma\gamma}}

\newcommand{\ttop}      {{t\bar{t}}}

\newcommand{\hsm}{{h_{\rm SM}}}

\renewcommand\epsilon{\varepsilon}

\newcommand{\dts}      {\Delta_S}

\title{
Radiative decays of a singlet scalar boson\\
through vector-like quarks}

\author[a]{Yeo Woong Yoon,}
\author[a,b,d]{Kingman Cheung,}
\author[c]{Sin Kyu Kang,}
\author[a]{and Jeonghyeon Song}

\affiliation[a]{Department of Physics,
Konkuk University, Seoul 143-701, Korea}
\affiliation[b]{Department of Physics, National Tsing Hua University,
Hsinchu, Taiwan 300}
\affiliation[c]{School of Liberal Arts, Seoul-Tech, Seoul 139-743, Korea}
\affiliation[d]{Physics Division, National Center for Theoretical Sciences,
Hsinchu, Taiwan 300}

\abstract{
If the standard model Higgs boson were much heavier,
it would appear as a broad resonance
since its decay into a pair of longitudinally polarized gauge bosons is
highly enhanced.
We study whether the same enhancement happens 
at loop level 
in a simple extension of the standard model with a singlet scalar boson $S$.
In order to focus on the loop effects,
we assume that $S$ does not interact with the standard model particles at tree level.
The singlet scalar $S$ is linked to the standard model world  
by vector-like quarks 
running in the loop.
We introduce three vector-like quark multiplets,
an $SU(2)_L$ doublet, an up-type singlet, and a down-type singlet.
There are two kinds of loop effects in the $S$ phenomenology, 
the mixing with the Higgs boson and the radiative decays
into $hh$, $WW$, $ZZ$, $gg$,
and $\gamma\gamma$ through the triangle loops.
We show that the crucial condition for enhancing loop effects
including the longitudinal polarization enhancement 
is
the large mass differences among vector-like quarks.
The current LHC constraints on $S$ from the heavy scalar searches 
and the Higgs precision data
are shown to be very significant: the mixing angle
with the Higgs boson should be smaller than about 0.1
for $m_S= 750$~GeV.
}

\begin{document}

\maketitle
\flushbottom
\section{Introduction}
The Standard Model (SM) has been more solidified by the Large Hadron Collider (LHC) data at 
$\sqrt{s}=13\tev$.
Supersymmetry models, composite Higgs models,
and other SM extensions are all strongly constrained.
Despite the absence of new signals, 
we believe that new physics beyond the SM must exist
since the issue of naturalness and the existence of dark matter
cannot  be explained within the SM.
There are two kinds of strategies to confront no signals,
pushing 
new particles out of the LHC reach~\cite{Kowalska:2016ent}
or introducing 
a hidden sector~\cite{Patt:2006fw,Strassler:2006im}.
If either is the case,
we turn to radiative corrections mediated by new particles
or the linking to the hidden sector~\cite{Englert:2011yb} in order to find
a clue about new physics.

An interesting question 
is how significantly the radiative correction or the linking changes tree level results.
They generally cause subleading corrections, but
there exist extreme cases also.
We may see entirely new signals
which are absent at tree level, such as the flavor changing neutral current processes through loops and 
the invisible Higgs decay modes through the mixing with a hidden sector.
In this work, we investigate a possibility that such new signals are so significant that
they could  be useful in the search for a new heavy scalar boson.

Among various tree level results,
the decay of a \emph{heavy} scalar boson shows an intriguing and unique feature,
the longitudinal polarization enhancement in its decay into a massive gauge boson pair.  
It is well-known that
 if the SM Higgs boson $h$ were much heavier, 
it would have decayed dominantly into a heavy gauge boson pair, $VV$ ($V=W,Z$).
The $\ttop$ channel, the next dominant one, has 
the maximum branching ratio about 19\%.
The extraordinarily large $\Gm(h \to VV)$ when $m_h \gg m_V$
is due to its decay into the longitudinal modes, $V_L V_L$~\cite{Cahn:1983ip,Kane:1984bb}. 
The longitudinal polarization vector of $V$ is proportional to
$p_V^\mu/m_V$ in the high energy limit, 
which leads to $\Gm(h \to V_L V_L) \propto m_h^3$.
The heavier the Higgs boson is, the larger the decay rate into $V_L V_L$ becomes.
For instance, 
$\Gm(h \to V_L V_L)$ is about 99\% of $\Gm(h \to VV)$ 
when $m_h \simeq 440\gev$.
Accordingly its total decay rate is also enhanced so that a heavy Higgs boson
becomes a broad resonance.

We wonder whether the same thing happens when a new heavy scalar boson  decays only radiatively.
To answer this question, we consider a simple extension of the SM
where
a  singlet scalar $S$~\cite{Schabinger:2005ei} 
and vector-like quarks (VLQs)~\cite{Ellis:2015oso} are introduced.
This can be considered as a simplified model.
Theoretically, a singlet scalar has drawn a lot of interest 
in the context of Higgs portal models \cite{Patt:2006fw}.
Its phenomenological signatures have been extensively studied~\cite{Barger:2007im,Dawson:2009yx,Fox:2011qc,Batell:2011pz,Bertolini:2012gu,Robens:2015gla}.
Heavy VLQs are also interesting as they
appear around the electroweak scale 
in many new physics models~\cite{Lavoura:1992np,Aguilar-Saavedra:2013qpa}.
The VLQs
are compatible with the current experimental results
while new heavy SM-like chiral quarks are excluded 
by the measurement of Higgs production rates~\cite{Anastasiou:2011qw,Anastasiou:2016cez}.
Moreover, the combination of a singlet scalar and vector-like quarks
is attractive:
it can shift 
the metastability of the electroweak vacuum in the SM~\cite{Cao:2013cfa,Xiao:2014kba,Batell:2012zw,Degrassi:2012ry,Buttazzo:2013uya};
it is crucial to construct a model
where all of the gauge and Yukawa couplings remain asymptotically safe 
up to infinite energy~\cite{Litim:2015iea,Pelaggi:2017wzr}.

In order to focus on the role of radiative corrections exclusively,
we consider a limiting scenario where $S$ does not couple to the SM particles at tree level.
The VLQs play the role of messengers
as connecting the SM particles and $S$ at loop level,
as interacting with the singlet scalar $S$, the SM Higgs boson, and the SM gauge bosons.
In order to allow the Yukawa interactions of VLQs with the Higgs boson,
we introduce three VLQ multiplets, an $SU(2)_L$ doublet, and two (up-type and down-type) 
$SU(2)_L$ singlets.
The presence of multiple VLQs shall be shown crucial in the $S$ phenomenology.
The VLQ loops have two kinds of implications.
First $S$ decays into $gg$, $\rr$, $WW$, $ZZ$, and $hh$ through triangle VLQ loops.
The singlet scalar $S$
can be produced and probed at high energy colliders.
Secondly,
$S$ is radiatively mixed with the Higgs boson.
Naive expectation is that the heavier the VLQs are, the smaller the loop corrections
become.
We shall show that this is not true.
Large mass differences in the VLQ mass spectrum
induce the longitudinal polarization enhancement
and increase the $S$-$h$ radiative mixing.
The obtained condition for 
the enhancement at loop level
shall help to study the physical properties of new
particles running in the loop.
These are our main results.



The paper is organized as follows.
In Sec.~\ref{sec:general},
we provide the general helicity amplitude framework for the decay of a scalar boson 
into a massive gauge boson pair and into a 
Higgs boson pair.
Section \ref{sec:model}
summarizes our new physics model with a singlet scalar $S$ and VLQs. 
The gauge and Yukawa couplings of the VLQs in terms of the mass eigenstates are given.
In Sec.~\ref{sec:loop},
we present our main analytic results of loop calculations.
The radiatively generated $S$-$h$ mixing and the loop induced
decay rates of $S \to hh, VV$ are to be shown.
In particular, the asymptotic behaviors of the loop functions
are very useful to understand the enhancement of $\Gm(S\to hh, VV)$
by large mass differences of the VLQs.
Section \ref{sec:results} is devoted to our numerical results
in a simple benchmark scenario.
The general physical properties of $S$ such as its branching ratio
and total decay rate are studied.
We also calculate the exclusion limits from the current LHC data 
of the heavy scalar searches and the Higgs precision observation.
The future prospect at the $13\tev$ LHC is also discussed. 
Section \ref{sec:conclusions} contains our conclusions.

\section{Decays of a scalar boson into $VV$ and $hh$}
\label{sec:general}

We consider a $\mathcal{J^{PC}}=0^{++}$
scalar particle $S$ which has a  mass $m_S$ and
a momentum $p^\mu$.
In the CP-conserving framework,
the most general coupling of $S$
to a pair of gauge bosons and that to 
a pair of the Higgs bosons can be parameterized by
\bea
\label{eq:def:ABC}
S(p) V_{\mu} (p_1) V^{\prime}_{\nu}(p_2) & :&  m_S
\left[ \mca \, g_{\mn} + \mcb \, \frac{p_{2\mu} p_{1\nu}}{m_S^2} \right],
\\ \nn
S h h & : & m_S \, \mcc,
\eea
where $\mca$, $\mcb$, and $\mcc$ are
dimensionless.

We write the helicity amplitudes for the decay $S\to V V'$
as
\bea
\langle V_{\mu} (p_1,\lm_1) V^{\prime}_{\nu}(p_2, \lm_2) |S(P) \rangle
&\equiv&
m_S \mct_{\lm_1 \lm_2},
\eea
where $\lm_{1}$ and $\lm_2$ are the helicities of the outgoing gauge bosons.
The dimensionless amplitudes $\mct_{\lm_1 \lm_2}$'s are then
written in terms of $\mca$ and $\mcb$ in Eq.~(\ref{eq:def:ABC}) as~\cite{Choi:2002jk}
\bea
\label{eq:hel:amp}
\mct_{++} &=&  \mct_{--} = -\mca,
\\ \nn
\mct_{00} &=&
\left\{
\begin{array}{ll}
\dfrac{m_S^2}{4 m_V^2}(2 \mca + \mcb) - (\mca+\mcb), &
\hbox{ if } m_V \equiv m_{V_1}=m_{V_2}\neq 0;
\\[5pt]
0, & \hbox{ if } m_{V_1}=0 \hbox{ or } m_{V_2}= 0,
\end{array}
\right.
\eea
and the other helicity amplitudes are zero.
The partial decay rates are
\bea
\label{eq:decay:rate}
\Gm (S \to VV') &=& \frac{1}{\mathcal{S}}\frac{\bt_{VV'}}{16 \pi } m_S
\sum_{\lm_1,\lm_2} \left| \mct_{\lm_1\lm_2}\right|^2,
\\[3pt] \nn
\Gm (S \to hh) &=&\frac{\bt_{hh}}{32\pi } m_S \left| \mcc \right|^2,
\eea
where the symmetric factor $\mathcal{S}$ is $1/2$ for two identical outgoing particles,
and
$\bt_{ij} = \sqrt{1 -  2(m_i^2+m_j^2)/m_S^2 + (m_i^2-m_j^2)^2/m_S^4}$.

When a scalar particle is heavy enough,
its decay into a massive gauge boson pair $VV$ ($V=W^\pm,Z$)
has a special feature.
The condition $m_S \gg m_V$ makes $\mct_{00}$
greatly enhanced if $(2\mca + \mcb) \neq 0$.
The SM Higgs boson, if its mass is greater than $2m_V$,
has
\bea
\label{eq:A:B:SM}
\mca^{h_{\rm SM}} =
\frac{2m_V^2}{v m_h},
\quad \mcb^{h_{\rm SM}} =0.
\eea
The partial
decay rate of $h_{\rm SM} \to V_L V_L$ is proportional to the cube of $m_h$
while that of $h_{\rm SM} \to V_T V_T$ is inversely proportional to $m_h$.
The heavier the Higgs boson is, the more dominant $h \to V_L V_L$
will become.
Another significant decay rate $\Gm(h\to \ttop)$ is linearly proportional to $m_h$.
The Higgs boson decay into $ V_L V_L$ is dominant in the total decay rate.
This is called the longitudinal polarization enhancement.

The partial decay rate of $S$ into a pair of SM Higgs bosons
is sizable if $\mcc \sim \mco (1)$.
In the MSSM,  an obvious scalar boson candidate which decays into $hh$ is
the heavy CP-even Higgs boson $H$.
However,
the decay into a pair of light Higgs bosons is suppressed in the alignment limit
since $\mcc$  is ~\cite{Craig:2013hca}
\bea
\label{eq:C:MSSM}
\mcc^{H_{\rm MSSM}} &=&
-
\frac{3 g_Z^2 \sin{4\bt}}{8} \frac{v}{M_H}
\left[
1+\mco (\cos(\bt-\al))
\right],
\eea
where 
$g_Z = g/\cos\theta_W$ and $\theta_W$ is the weak mixing angle.
The partial decay rate $\Gm(H \to hh)$ is inversely proportional to the
heavy Higgs mass: there is no
enhancement in the $hh$ decay channel.

\section{Model with a singlet scalar and vector-like quarks}
\label{sec:model}

We consider a simple extension of the SM
by introducing a CP-even singlet scalar boson $S_0$,
a VLQ doublet $\mcq_{L/R}$, two
VLQ singlets $\mcu_{L/R}$ and $\mcd_{L/R}$:
\begin{eqnarray}
\mcq_{L/R}=\left( \begin{array}{c}
              \mcu^{\prime} \\
              \mcd^{\prime} \end{array} \right)_{L/R},
\quad
\mcu_{L/R}\,, \quad \mcd_{L,R}.
\end{eqnarray}
The $SU(3)_c \times SU(2)_L \times U(1)_Y$ quantum numbers of
$\mcq_{L/R}$, $\mcu_{L/R}$, $D_{L/R}$ are
$(\mathbf{3}, \mathbf{2}, 1/3)$,  $(\mathbf{3}, \mathbf{1},4/3)$,
and $ (\mathbf{3}, \mathbf{1},  -2/3)$, respectively.
The hypercharges of VLQs are set to be the same as the SM quarks.
Different assignment
shall affect the decays of $S$ into $ZZ$ and $\rr$.


The most general scalar potential of the SM Higgs doublet $H$
and a real singlet scalar $S_0$
is \cite{Chen:2014ask}
\bea
\label{eq:V:HS}
V(H,S_0) &=& -\mu^2 H^\dagger H +\lm( H^\dagger H)^2
+\frac{a_1}{2} S_0 H^\dagger H + \frac{a_2}{2} S_0^2 H^\dagger H
\\ \nn
&& + b_1   S_0 + \frac{b_2}{2}  S_0^2 + \frac{b_3}{3} S_0^3+ \frac{b_4}{4} S_0^4.
\eea
Note that we do not assume any discrete $Z_2$ symmetry for $S_0$.
When defining the neutral component of $H$ as
$\phi_0 = (v_0 + h_0)/\sqrt{2}$ and
the VEV of the singlet field as
$\langle S_0 \rangle = x$, the extrema of the potential satisfy
\bea
\label{Eq:min:condition}
\frac{\rd V(v_0/\sqrt{2},x)}{\rd v_0} =0, \quad \frac{\rd V(v_0/\sqrt{2},x)}{\rd x} =0.
\eea
Although there exist many possible extrema, the
true minimum of $H$ should generate proper EWSB, i.e., $v_0=v=246\gev$.
On the other hand, the VEV of $S$
is free to choose
since the shift of the singlet field, $S \to S +\dts$,
corresponds to redefining the parameters of $a_{1,2}$ and $b_{1,\cdots,4}$.
There is no change in physics.
Without loss of generality we take $(v_0,x) = (v, 0)$.
Note that the choice of vanishing VEV for $S_0$ eliminates the tadpole
term of $S_0$.
The minimization conditions in Eq.~(\ref{Eq:min:condition}) become
\bea
\label{eq:min:condition:0Svev}
\mu^2 &=& \lm v^2,
\quad
b_1 = -\frac{v^2}{4}a_1.
\eea

The Yukawa terms of VLQs with the singlet $S_0$ and the SM Higgs doublet $H$
as well as their mass terms are
\bea
\label{eq:L:Yukawa}
-\lg_Y&=&
S_0 \left[
y_\mcq \bar{\mcq}\mcq + y_\mcu \bar{\mcu}\mcu + y_\mcd \bar{\mcd}\mcd
\right]
+M_\mcq  \bar{\mcq}\mcq + M_\mcu \bar{\mcu}\mcu + M_\mcd \bar{\mcd}\mcd
\\ \nn &&
+\left[
Y_\mcd \bar{\mcq}_L H \mcd_R + Y'_\mcd \bar{\mcq}_R H \mcd_L
+
Y_\mcu \bar{\mcq}_L \widetilde{H} \mcu_R
+
Y'_\mcu \bar{\mcq}_R \widetilde{H} \mcu_L + H.c.
\right],
\eea
where $\widetilde{H} = i \tau_2 H^*$.
For simplicity, we assume
$y_\mcq = y_\mcu = y_\mcd \equiv y_S$, $Y_{\mcu}=Y_{\mcu'}$, and $Y_{\mcd}=Y_{\mcd'}$
in what follows.

The VLQ mass matrix
$\mathbb{M}_{F}$  in the basis of $(F', F)$ where $F=\mcu,\mcd$
is
\bea
\mathbb{M}_F =
\left(
\begin{array}{cc}
~M_\mcq~ & ~\frac{Y_{F} v}{\sqrt{2}}~ \\[5pt] \frac{Y_{F} v}{\sqrt{2}} & M_F
\end{array}
\right),
\eea
which is diagonalized by
the mixing matrix of
\bea
\label{eq:mixing:R}
\mathbb{R}_{\theta_F}
=
\left(
\begin{array}{cc}
\cthf & -\sthf \\ \sthf & \cthf
\end{array}
\right).
\eea
Here we adopt simplifying notations of $c_x =\cos x$ and $s_x =\sin x$.
The $Y_\mcu$ and $Y_\mcd$ terms
generate the mixings between VLQ doublet and VLQ singlets.
If $Y_F \sim \mco(1)$,
VLQ mixing angles are small
since VLQs are expected to be heavy.
The mass eigenvalues and the mixing angle  are then
\bea
\label{eq:MF:sinF}
M_{{F_1},{F_2}}&=& \frac{1}{2} \left[M_\mcq+M_F \mp
\sqrt{ ( M_F-M_\mcq )^2+ 2 Y_F^2 v^2 }
\right],
\\ \nn
\stthf &=&  \,\frac{ \sqrt{2} Y_F v }{M_{F_2} - M_{F_1}},
\eea
where $M_{F_1}<M_{F_2}$.

The Yukawa terms of the VLQ mass eigenstates become
\bea
-\lg_{\rm Yukawa}
&=& y_S S_0 \sum_{i}
\left[
\overline{\mcu}_i \mcu_i + \overline{\mcd}_i \mcd_i
\right]
+
h_0 \sum_{F} \sum_{i,j}
 y_{_{h F_i F_j} }
\bar{F}_i F_j
,
\eea
where $F=\mcu,\mcd$, $i,j=1,2$, and
$y_{_{h F_i F_j} }$ are
\bea
\label{eq:y:hFF}
y_{_{h F_1 F_1}} &=& - y_{_{h F_2 F_2}}
=  -\frac{Y_F}{\sqrt{2}}\,s_{2\theta_F},
\quad
y_{_{h F_1F_2}} = y_{_{h F_2 F_1}} = -\frac{Y_F}{\sqrt{2}}\,c_{2 \theta_F}.
\eea

The gauge interaction Lagrangian in terms of the VLQ mass eigenstates is
\bea
\lg_{\rm gauge} &=&
e  A_\mu \sum_F \sum_{i} Q_F \bar{F}_i \gm^\mu F_i
+ g_Z Z_\mu \sum_{F} \sum_{i,j} \gh_{_{Z F_i F_j}}
\bar{F}_i \gm^\mu F_j
\\ \nn
&& +
\frac{g}{\sqrt{2}}\,
\Big[
W^{+\mu} \sum_{i,j} \gh_{_{W \mcu_i \mcd_j}}
\bar{\mcu}_i \gm_\mu \mcd_j
+ H.c. \Big].
\eea
Here $Q_F$ is the electric charge of the fermion $F$
and the effective gauge couplings $\gh_{_{VFF'}}$ are
\bea
\label{eq:gh:VFF}
\gh_{_{Z F_1 F_1 }} &=& \bar{g}_{\mcq}^{v} \cthf^2 + \bar{g}_{F}^{v} \sthf^2, \quad
\gh_{_{ZF_2F_2}}= \bar{g}_{\mcq}^{v}\sthf^2 + \bar{g}_{ F}^v \cthf^2,
\\ \nn
\gh_{_{Z F_1 F_2}} &=&
\left( \bar{g}_{\mcq}^{v} - \bar{g}_{F}^{v} \right) \sthf\cthf,
\\ \nn
\gh_{_{W\mcu_1 \mcd_1}} &=&
\cthu \cthd ,\quad 
\gh_{_{W \mcu_1 \mcd_2}} =
 \cthu \sthd, 
\\ \nn
\gh_{_{W \mcu_2 \mcd_1}} &=&
\sthu \cthd , \quad
\gh_{_{W \mcu_2 \mcd_2} } =
  \sthu\sthd ,
\eea
where $\gh_{_{VFF'}}=\gh_{_{VF'F}}$ and 
$\bar{g}_{\mcf}^v = \tfrac{1}{2}T_3^{\mcf} - s_{W}^2 Q_\mcf$ 
for $\mcf=\mcq,\mcu,\mcd$.
There is a big difference between $h$-$F$-$F'$ couplings and $V$-$F$-$F'$ couplings.
In the limit of $\theta_{\mcu,\mcd}\ll 1$,
the gauge couplings to different mass eigenstates of VLQs
(e.g. $\gh_{_{V F_1 F'_2}}$)
are suppressed by $s_{\theta_F}$.
On the contrary, the VLQ couplings to the Higgs boson
are suppressed for the same mass eigenstates.

Without the $Z_2$ symmetry,
the $S_0$ field can couple to the SM particles at tree level.
Since the singlet scalar $S_0$ is neutral under all quantum numbers of the SM
gauge group,
the only possible renormalizable couplings of $S_0$ to the SM particles
at tree level
are to the Higgs boson
through $a_1$ and $a_2$ terms in Eq.~(\ref{eq:V:HS}).
However, a nonvanishing $a_1$ term
will generate the $S$-$h$ mixing with the mixing angle $\eta$,
which shall change the Higgs coupling modifiers of $\kp_V$ and $\kp_f$ into $c_\eta$.
According to the global fit analysis of the LHC Higgs precision data~\cite{Dolan:2016eki,kappa:2015,Cheung:2015dta},
$c_\eta$ is very close to 1.
Nonzero $a_1$ builds up 
some tension with the Higgs boson constraints.
Moreover, our main question is whether the unique characteristic of a heavy scalar boson
such as the longitudinal polarization enhancement
remains even at loop level.
Therefore,
we consider a limiting scenario in which
the singlet scalar
has no tree level couplings with the Higgs boson:
\bea
\label{eq:no:tree}
a_1^{\rm tree}=0=a_2^{\rm tree}.
\eea

\section{The effects of the VLQ loops}
\label{sec:loop}
In the previous section,
we suggested a rather  extreme scenario
where $S_0$
does not interact with the SM Higgs boson at tree level.
The singlet field $S_0$ could be considered as a field in a hidden sector.
In the model,
the visible sector and the hidden sector are connected via VLQ loops:
the VLQs play the role of messengers.
There are two phenomenological implications: (i)
the singlet-Higgs mixing and (ii)
the radiative decays of $S$  into SM particles.
We study the effects at one loop level.

\subsection{$S$-$h$ mixing and Higgs Modifiers}

\begin{figure}[h] \centering
\begin{center}
\includegraphics[width=.35\textwidth]{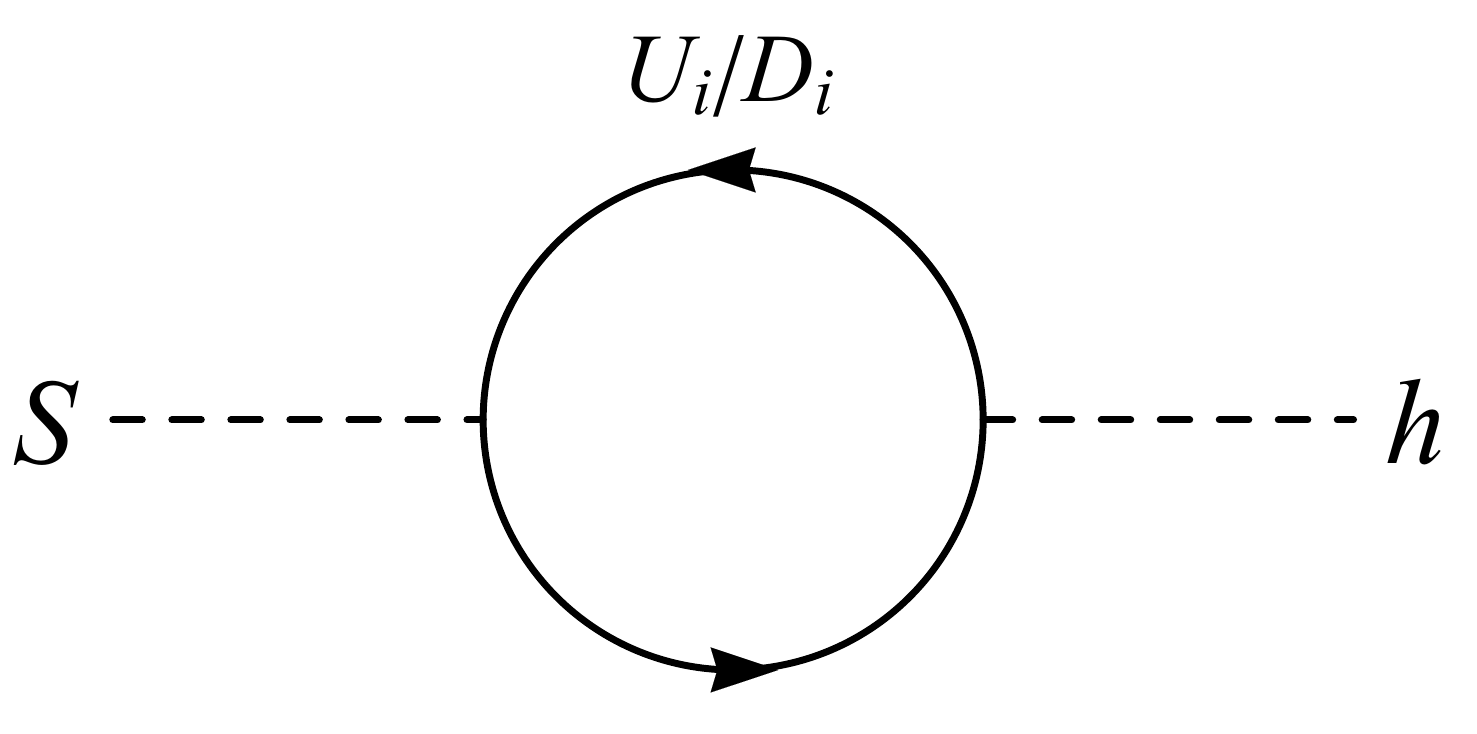}
\end{center}
\caption{\label{fig-Feyn-Sh-mixing}
\baselineskip 3.0ex
Feynman diagrams for the loop induced $S$-$h$ mixing.
}
\end{figure}

First, the $S$-$h$ mixing is radiatively generated through
the VLQ loops as shown
in Fig.~\ref{fig-Feyn-Sh-mixing}.
The scalar-mass-squared matrix in the basis
of $(h_0,S_0)$
becomes
\bea
\mathbb{M}_{hS}^2
\equiv
\left(
\begin{array}{cc}
2 \lm v^2 & \dt M_{Sh}^2 \\
\dt M_{Sh}^2 & M_{SS}^2
\end{array}
\right)\,,
\eea
where $M_{SS}^2= b_2$
since we have used the conditions in
Eq.~(\ref{eq:min:condition:0Svev})
for our choice of the vacuum $(v_0,x) = (v,0)$.
At one loop level, we have
\bea
\dt M_{Sh}^2 &=& -
\frac{y_S N_c}{4\pi^2}
\sum_F \sum_i
y_{_{hF_i F_i }} M_{F_i}^2
\bigg[4(\tau^S_{F_i}-1)g(\tau^S_{F_i}) -4\tau^S_{F_i}+5
\bigg]\,,
\eea
where $N_c=3$ is the color factor of the VLQ,
$F=\mcu,\mcd$, $i=1,2$,
$\tau^i_{j} = m_i^2/(4m_j^2)$,  and
$y_{_{h F F'}}$'s are in Eq.~(\ref{eq:y:hFF})\footnote{
There is UV divergence in the one loop calculation of $\delta M_{Sh}^2$ which must be properly renormalized. Detailed description on the renormalization of the whole model is in preparation~\cite{future}. 
}.
The loop function $g(\tau)$ is given by
\bea
g(\tau) &=& \left\{ \begin{array}{ll}
\sqrt{\tau^{-1}-1}\arcsin\sqrt{\tau} & \hbox{ if } \tau \le 1; \\[3pt]
\frac{\sqrt{1-\tau^{-1}}}{2}\left[
\log\frac{1+\sqrt{1-\tau^{-1}}}{1-\sqrt{1-\tau^{-1}}}-i\pi \right]
& \hbox{ if } \tau > 1\,. \end{array} \right.
\eea
Note that $\dt M_{Sh}^2$ vanishes if $M_{F_1}=M_{F_2}$ 
since $y_{h F_1 F_1} = - y_{h F_2 F_2}$: see Eq.~(\ref{eq:y:hFF}).
Significant $S$-$h$ mixing requires sizable mass differences of $F_1$ and $F_2$.

The mass eigenvalues and the $S$-$h$ mixing angle $\eta$ are
\bea
m_{h,S}^2&=&
\frac{1}{2}
\left(
M_{hh}^2+M_{SS}^2 \mp \sqrt{(M_{SS}^2-M_{hh}^2)^2+ 4 (\dt M_{Sh}^2)^2 }
\right),
\\ \nn
s_{2\eta} &=& -
\frac{ 2 \,\dt M_{Sh}^2}{m_S^2-m_h^2 }
,
\eea
where we use the $S$-$h$ mixing matrix $\mathbb{R}_\eta$
in Eq.~(\ref{eq:mixing:R}).
Since $\dt M_{Sh}^2$ is radiatively generated,
we expect $s_\eta \ll 1$.
We take the mass eigenstate $h = c_\eta h_0 - s_\eta S_0$
to be the observed Higgs boson with a mass of $125\gev$, and assume
$S$ to be heavy such as  $m_S \gsim 500\gev$.


The nonzero $S$-$h$ mixing changes
the Higgs coupling modifiers
of $\kp_Z$, $\kp_W$, $\kp_t$, $\kp_\tau$, and $\kp_b$
to be $c_\eta$.
The loop induced decays of the Higgs boson into $gg$ and $\rr$
have additional loop contributions from VLQs.
We define $\kp_g$ and $\kp_\gm$ as
\bea
\lg_{\rm Higgs}
&=& \kp_g c_g^\sm
\,\frac{h}{v}\, G^{a\mn} G^a_\mn
+ \kp_\gm c_\gm^\sm
\,\frac{h}{v}\, F^{\mn} F_\mn\,.
\eea
The SM values $c_g^\sm$ and $c_\gm^\sm$ are
\bea
c_g^\sm \equiv
\frac{\alpha_s}{16\pi} 
 A^\sm_{hgg}
, \quad
c_\gm^\sm \equiv
\frac{\alpha_e}{8 \pi}
 A^\sm_{h\rr},
\eea
where $A^\sm_{hgg} = \sum_{f=t,b} A_{1/2}(\tau^h_f)$,
$A^\sm_{h\rr}
=
A_1(\tau^h_W) +
\sum_{f=t,b,\tau} N_C^f Q_f^2 A_{1/2}(\tau^h_f)$, and $ A_{1/2}(\tau)$ and $A_1(\tau)$
are referred to Ref.~\cite{Djouadi:2005gi}.
The modifiers
$\kp_g$ and $\kp_\gm$ receive two kinds of new contributions.
One is from the modified couplings of $h$ to the SM particles through the $S$-$h$ mixing.
The other is from the triangle VLQ loops, parameterized by
\bea
\label{eq:Ahgg}
\mathcal{A}^{\rm VLQ}_{hgg}
&=&
\sum_{F}\sum_i
y_{_{h F_i F_i}}  \frac{v}{M_{F_i}} A_{1/2}(\tau^h_{F_i})
,
\\ \nn
\mathcal{A}^{\rm VLQ}_{h\rr}
&=&
\sum_{F}\sum_i
N_C
Q_{F_i}^2 y_{_{h F_i F_i}}
\frac{v}{M_{F_i}} A_{1/2}(\tau^h_{F_i})
,
\eea
where $F=\mcu,\mcd$,
$i=1,2$,
$\tau^i_j=m_i^2/(4 m_j^2)$.
Then $\kp_g$ and $\kp_\gm$ are
\bea
\kp_{g,\gm} &=&
\frac{ c_\eta A^\sm_{hgg,h\rr}
+
\mathcal{A}^{\rm VLQ}_{hgg,h\rr} }
{A^\sm_{hgg,h\rr}
}
.
\eea
Since $y_{_{h F_1 F_1}}$ and $y_{_{h F_2 F_2}}$
in Eq.~(\ref{eq:y:hFF}) have opposite signs,
both $\dt M_{Sh}^2$ as well as $\mathcal{A}^{\rm VLQ}_{hgg,h\rr}$ are suppressed
when $M_{F_1} \simeq M_{F_2}$.

\begin{figure*}[t!]
    \centering
    \begin{subfigure}[b]{0.5\textwidth}
        \centering
        \includegraphics[width=0.8\textwidth]{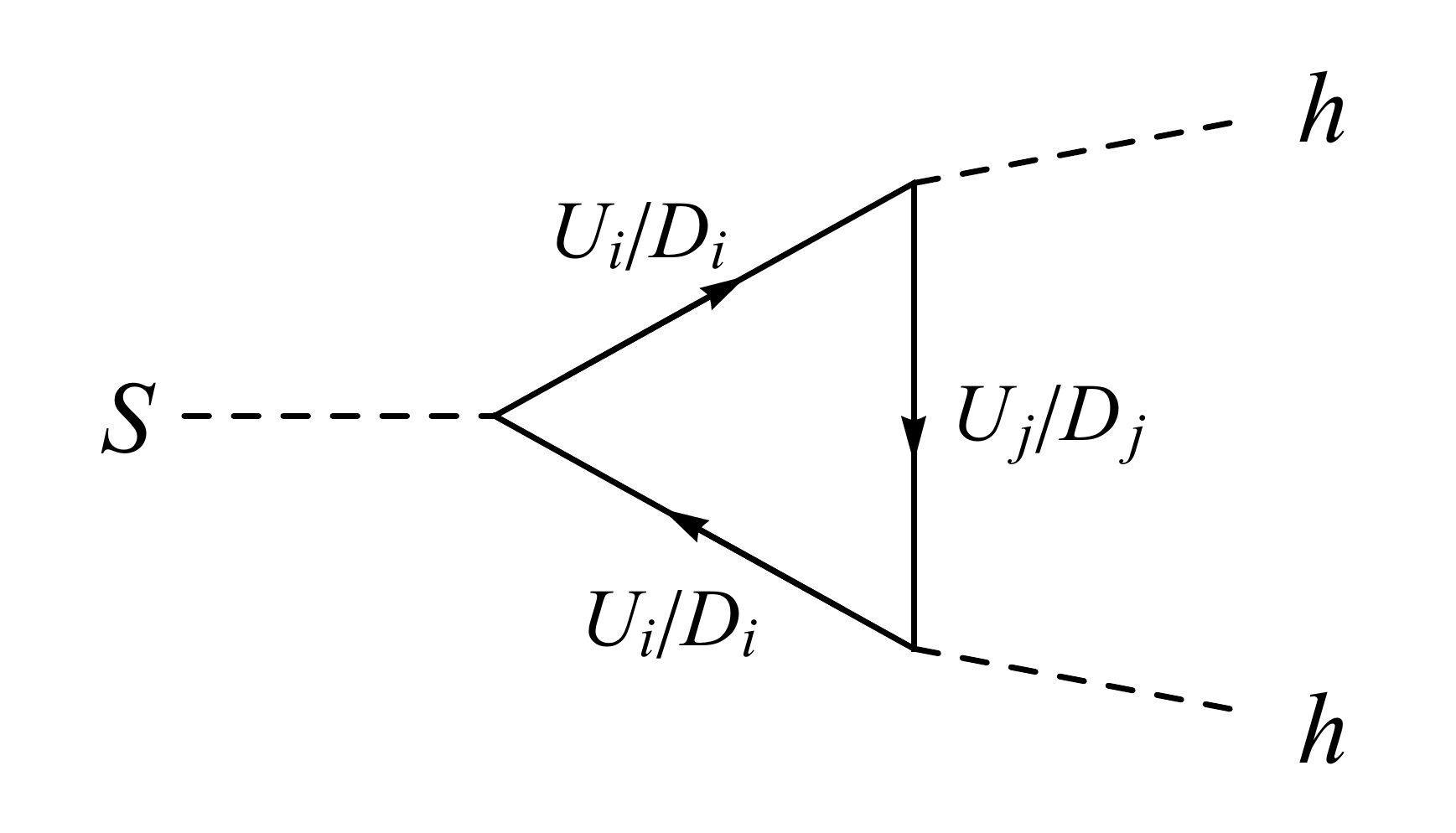}
        \caption{}
    \end{subfigure}%
    ~
    \begin{subfigure}[b]{0.5\textwidth}
        \centering
        \includegraphics[width=0.8\textwidth]{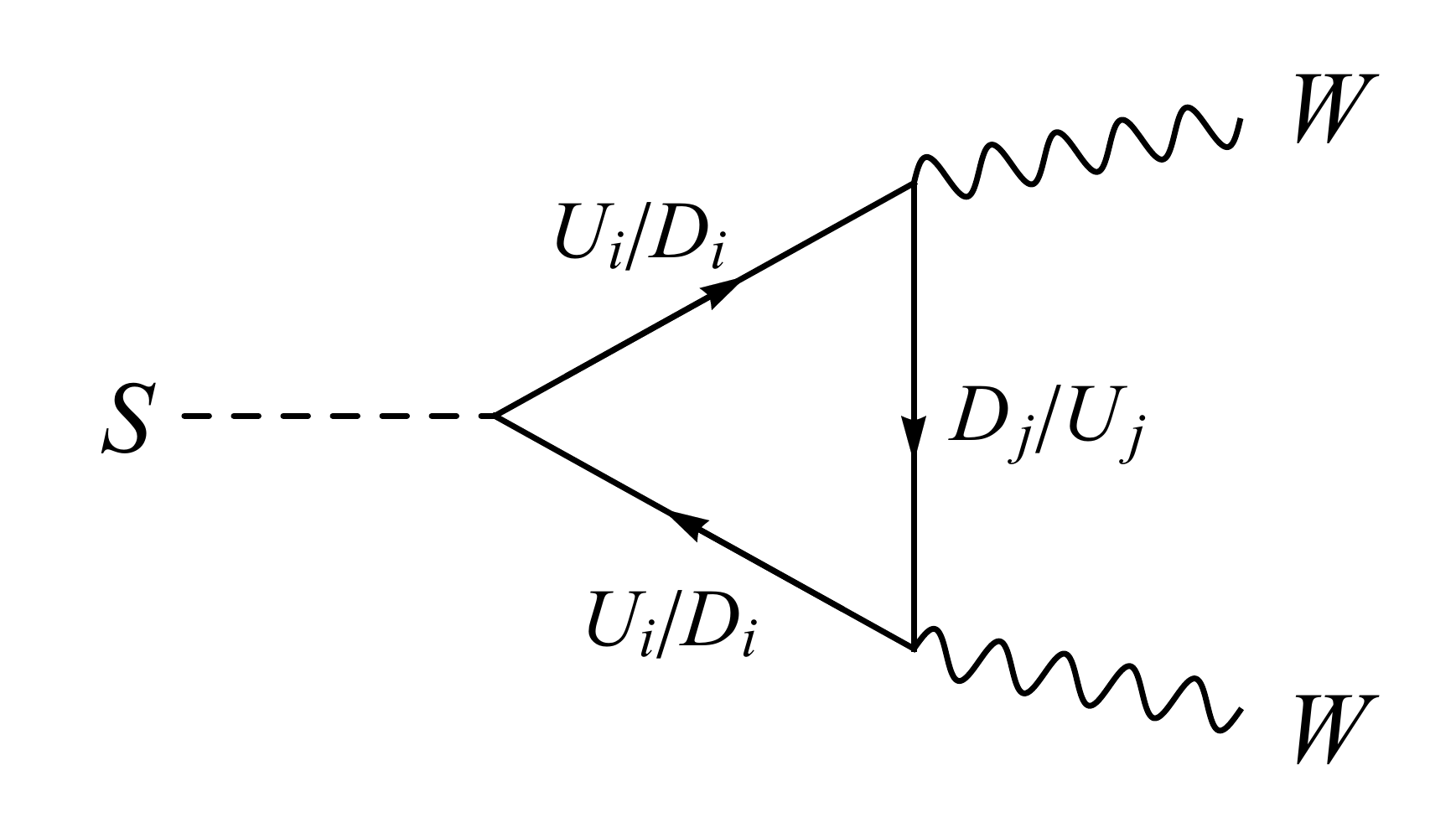}
        \caption{}
    \end{subfigure}
    \caption{\label{fig-Feyn-SWW-Shh}
Feynman diagrams of  $S \to hh$ and $S \to WW$
from the VLQ loops.
}
\end{figure*}

\subsection{Radiative Decays of $S$}

Another important effect of the VLQ loops
is the radiative decay of $S$
into the SM particles,
which occurs through the $S$-$h$ mixing as in Fig.~\ref{fig-Feyn-Sh-mixing}
and/or through the triangle VLQ loops into a gauge boson pair
or a Higgs boson pair as in Fig.~\ref{fig-Feyn-SWW-Shh}.
Since we consider the case of $m_S \gsim 500\gev$,
the main decay modes are into $\ttop$, $gg$, $\rr$, $WW$,  $ZZ$, and $hh$.

The decay of $S$ into a top quark pair is only through the $S$-$h$ mixing.
The partial decay rate is
\bea
\Gm (S \to \ttop ) = s_\eta^2 \, \Gm (\hsm\to \ttop) \Big|_{m_\hsm = m_S}.
\eea
%
Another important decay channel is $S \to hh$
shown in Fig.~\ref{fig-Feyn-SWW-Shh}(a).
The vertex $\mcc$
in Eq.~(\ref{eq:def:ABC}) at one loop level
is
\bea
\label{eq:Ctot}
\mcc &=&  \frac{y_S N_c}{4\pi^2}\sum_{F} \;\sum_{i,j}
y_{_{h F_i F_j}}^2  \,
\mcc_T(m_h,m_S, M_{F_i}, M_{F_j})
 + \frac{3 m_h^2}{v m_S} s_\eta
 \, ,
\eea
where $y_{_{hFF^\prime}}$
are given in Eq.~(\ref{eq:y:hFF}).
The first term is due to the triangle diagrams while the second one is from the $S$-$h$ mixing.
The asymptotic expression\footnote{Full expressions of form factors such as
$\mca_T$, $\mcb_T$, and $\mcc_T$
are to be presented in Ref.~\cite{future}.} of
$\mcc_T(m_h,m_S,M_{F_i},M_{F_j})$
when $m_h \ll m_S$ and $\Dt_F
\ll M_\mcf$, where $\Dt_F
= M_{F_j}-M_{F_i}$ and 
$M_\mcf =
(M_{F_i} + M_{F_j})/2)
$,
is very useful to understand the enhancement of $\Gm (S\to hh)$ in some parameter space:
\begin{eqnarray}
\label{eq:CM:asymptotic}
\sqrt{\tau}\,{\cal C}_T &=& 2 + (1-2\tau^{-1})f(\tau) -2g(\tau)
\nn\\
&& ~
+\Big(\frac{\Delta_F^2}{M_{\cal F}^2}\Big)\bigg[
\frac{8\tau^2+49\tau-48}{12\tau(1-\tau)} + \frac{(\tau^3+12\tau^2-26\tau+16)g(\tau)}{4\tau(1-\tau)^2}
\bigg]\nn\\
&& ~
+\Big(\frac{m_h^2}{m_S^2}\Big)
\Big[\frac{2(6-\tau)}{3}+\frac{2(\tau-2)f(\tau)}{\tau}\Big]
+{\cal O}\Big(\frac{\Delta_F^4}{M_{\cal F}^4}\Big)
+{\cal O}\Big(\frac{m_h^4}{m_S^4}\Big)
\,.
\end{eqnarray}
where $\tau =m_S^2/(4 M_\mcf^2)$ and $f(\tau)$ is referred to Ref.~\cite{Djouadi:2005gi}.
Note that the odd power terms in $(\Delta_F/M_{\cal F})$ are 
neglected since they cancel each other after the summation in Eq. (\ref{eq:Ctot}).
If $y_S, Y_{\mcu,\mcd} \sim \mco(1)$, 
$\mcc$ is not suppressed by large $m_S$,
contrary to the case of a heavy CP-even scalar $H$ of the MSSM in Eq.~(\ref{eq:C:MSSM}).
Another important result is that
the partial decay rate $\Gm(S\to hh)$ increases with $\Dt_F$,
the mass difference between
$M_{F_i}$ and $M_{F_j}$.
Since $\Dt_F$ is proportional to the Higgs VEV from the SM-like Yukawa couplings of the VLQs to the Higgs boson,
the enhancement of $S \to h h $ can be considered as non-decoupling effects.

The VLQ loops also allow the decay of $S$
into a massive gauge boson pair $VV$ ($V=W,Z$)
as shown in Fig.~\ref{fig-Feyn-SWW-Shh}(b).
The dimensionless parameters $\mca$ and $\mcb$ 
in Eq.~(\ref{eq:def:ABC}) are
\begin{eqnarray}
\nn
{\cal A}_{WW} &=&
\frac{g^2 y_S N_c }{8\pi^2}\sum_{i,j}
\left[
   \gh_{_{W\mcu_i \mcd_j}}^2
{\cal A}_T(m_W,m_S, M_{\mcu_i}, M_{\mcd_j})
+ \{ \mcu \leftrightarrow \mcd \}
\right] +
\frac{2 m_W^2}{v m_S} s_\eta
, \\ \nn
 {\cal B}_{WW} &=& \frac{g^2 y_S N_c}{8\pi^2}\sum_{i,j}
\left[
  \gh_{_{W\mcu_i \mcd_j}}^2
{\cal B}_T(m_W,m_S, M_{\mcu_i}, M_{\mcd_j})
+ \{ \mcu \leftrightarrow \mcd \}
\right]\,,
\\ \nn
{\cal A}_{ZZ} &=&
 \frac{g_Z^2 y_S N_c}{4\pi^2}\sum_{i,j}\left[
  \gh_{_{Z {\cal U}_i {\cal U}_j}}^2
 {\cal A}_T(m_Z,m_S, M_{{\cal U}_i}, M_{{\cal U}_j})
 + \{ \mcu \leftrightarrow \mcd \}
\right]\,
 +\frac{ 2m_Z^2}{v m_S} s_\eta
,\\ \label{eq:AVV:BVV}
{\cal B}_{ZZ} &=&
\frac{g_Z^2 y_S N_c}{4\pi^2}\sum_{i,j}
\left[
 \gh_{_{Z{\cal U}_i {\cal U}_j}}^2\,
 {\cal B}_T(m_Z,m_S, M_{{\cal U}_i}, M_{{\cal U}_j})\,
 + \{ \mcu \leftrightarrow \mcd \}
\right]\,,
\end{eqnarray}
where
$i,j=1,2$.
$\mca_{VV}$ consists of two parts,
one from the triangle VLQ loops and the other from the $S$-$h$ mixing,
while
$\mcb_{VV}$ is only from the triangle loops.

Our main question is whether the longitudinal polarization enhancement
in $S\to VV$ remains significant at loop level, which happens
when $2\mca + \mcb \neq 0$ as shown in Eq.~(\ref{eq:hel:amp}).
The $S$-$h$ mixing induced terms,
proportional to $s_\eta$ in Eq.~(\ref{eq:AVV:BVV}),
appear only in $\mca$ and thus
generate the longitudinal polarization enhancement.
The condition that the triangle VLQ loops induce the enhancement 
is easy to see through
the asymptotic behaviors of $\mca_T$ and $\mcb_T$
in the limit of $\Dt_F \ll M_\mcf$
and $m_{V} \ll m_S$, given by
\begin{eqnarray}
\label{eq:AM}
\sqrt{\tau}\,{\cal A}_T &=& 1+(1-\tau^{-1}) f(\tau)
\\ \nn && ~
+\Big(\frac{\Delta_F^2}{M^2}\Big)\Big[
-\frac{1}{4} + \frac{(3\tau-4)f(\tau)}{4\tau^2}
+\frac{(\tau^2+4\tau-8)g(\tau)}{4\tau(\tau-1)}
\Big]
\\ \nn && ~
+2\Big(\frac{m_V^2}{m_S^2}\Big)
\Big[ 3-\tau -\tau^{-1} f(\tau) - 2g(\tau) \Big]
 +{\cal O}\Big(\frac{\Delta_F^4}{M^4}\Big)
 +{\cal O}\Big(\frac{m_V^4}{m_S^4}\Big)
\,,
\\
\label{eq:BM}
\sqrt{\tau}\,{\cal B}_T &=& -2-2(1-\tau^{-1}) f(\tau)
\\ \nn && ~
+\Big(\frac{\Delta_F^2}{M^2}\Big)\bigg[
\frac{5}{2} + \frac{(8-5\tau)f(\tau)}{2\tau^2}
-\frac{(\tau^2+12\tau-16)g(\tau)}{2\tau(\tau-1)}
\bigg]\\ \nn
&& ~
+4\Big(\frac{m_V^2}{m_S^2}\Big)
\Big[\tau-4 +(2-\tau)\tau^{-1} f(\tau) + 2g(\tau) \Big]
+{\cal O}\Big(\frac{\Delta_F^4}{M^4}\Big)
+{\cal O}\Big(\frac{m_V^4}{m_S^4}\Big)\,,
\end{eqnarray}
where $\tau =m_S^2/(4 M_\mcf^2)$.
Equations (\ref{eq:AM}) and (\ref{eq:BM}) show
that $2\mca + \mcb \sim \mco\left(m_V^2/m_S^2 \right) $ if $\Delta_F=0$.
Sizable mass differences of VLQs
are crucial for the longitudinal polarization enhancement
through the triangle VLQ loops.

The last category of the radiative decays of $S$ is into $gg$, $\rr$, and
$Z\gm$.
When at least one of the outgoing gauge bosons is massless,
there is no longitudinal polarization mode as shown in Eq.~(\ref{eq:hel:amp}).
The
$\mca$'s are
\bea
{\cal A}_{\gamma\gamma} &=& \frac{ e^2 y_S N_c}{4\pi^2}
\sum_{F} \sum_i Q_{F_i}^2 \frac{1}{\sqrt{\tau_{F_i}}}
\left[1+(1-\tau_{F_i}^{-1})f(\tau_{F_i}) \right],
\\ \nn
{\cal A}_{gg} &=& \
\delta^{ab}
\frac{ g_s^2 y_S }{8\pi^2}
\sum_{F} \sum_i
\frac{1}{\sqrt{\tau_{F_i}}}
\left[
1+(1-\tau_{F_i}^{-1})f(\tau_{F_i}) \right]\,,
\\ \nn
{\cal A}_{Z\gm} &=&
\frac{ e \, g_Z\, y_S N_c}{2\pi^2}
\sum_F \sum_i Q_{F_i} \gh_{_{Z F_i F_i}}
\frac{1}{\sqrt{\tau_{F_i}}}
\left[
-1-(1-\tau_{F_i}^{-1})f(\tau_{F_i})
+ \mco \left( \frac{m_Z^2}{m_S^2} \right) \right],
\eea
where
$a,b$ are color indices of the outgoing gluons,
$F=\mcu,\mcd$, $i=1,2$, and $\tau_{F_i} = m_S^2/(4 M_{F_i}^2)$.
The ${\cal B}$'s can be obtained by using Ward identity as follows
\begin{eqnarray}
{\cal B}_{\rr} = -2 {\cal A}_{\rr},~~
{\cal B}_{gg} = -2 {\cal A}_{gg},~~
{\cal B}_{Z\gamma} = -2 \Big( 1- \frac{m_Z^2}{m_S^2}\Big)^{-1}{\cal A}_{Z\gamma}\,.
\end{eqnarray}

The final comment in this section is the importance of the VLQ Yukawa couplings 
with the \emph{Higgs boson}
in enhancing the radiative decay rates of $S$. 
If we do not allow the $Y_\mcu$ and $Y_\mcd$ terms,
which happens for example when we introduce only one VLQ multiplet,
the $S$-$h$ mixing and the $S \to hh$ decay will be absent.
In addition, the VLQs running in the triangle VLQ loops for the decay of $S\to WW,ZZ$
have the same masses because of no VLQ mixing.
There is no longitudinal polarization enhancement
and thus the signal rates of the radiative decays have typical loop suppression~\cite{Cao:2009ah}.
In summary, the presence of the VLQ doublet \emph{and} the VLQ singlets
are crucial for the enhanced radiative decays of $S$.

\section{Numerical Results}
\label{sec:results}

The phenomenological characteristics of the singlet scalar $S$
depend on the model parameters of $y_S$, $m_S$,
$Y_{\mcu,\mcd}$,
$M_\mcq$,
$M_{\mcu}$ and $M_{\mcd}$.
The $y_S$ contributes equally to all
of the partial decay rates of $S$ by the common factor of $y_S^2$
since $S$ decays only radiatively through VLQ loops in our model.
The branching ratios of $S$ are independent of $y_S$.
The $m_S$ dependence on the branching ratios is also weak for the heavy $S$.
The $Y_{\mcu,\mcd}$,
$M_\mcq$, and
$M_{\mcu,\mcd}$
specify the VLQ mass matrices and thus the mass difference $\Dt_F$.
Since
$Y_{\mcu}$ and $Y_{\mcd}$ also quantify the VLQ couplings with the Higgs boson,
they are the most crucial parameters.

Therefore,  we consider a simple benchmark parameter line, given by
\bea
\label{eq:benchmark}
M_\mcq = M_\mcu=M_\mcd, \quad Y_\mcu=0, \quad Y_\mcd \hbox{ varies}.
\eea
We found that the results in this simple case
display the main characteristic features of the radiative decays of $S$.
The VLQ mass spectra become
\bea
\label{eq:mass:benchmark}
M_{\mcu_1}&=&M_{\mcu_2}=M_\mcq,\quad
M_{\mcd_{1,2}} = M_\mcq \mp \frac{1}{\sqrt{2}} |Y_\mcd| v.
\eea
Note that $\mcd_1$ becomes the lightest VLQ 
and $\Dt M_{\mcu_1 \mcd_1} = \Dt M_{\mcd_2 \mcu_1} = (1/2) \Dt M_{\mcd_2 \mcd_1}$
where $\Dt M_{ij} \equiv M_i - M_j $.
Our setting of $Y_\mcu \neq Y_\mcd$
generates a sizable mass difference $\Dt M_{\mcu_1 \mcd_1}$
which is essencial for the longitudinal polarization enhancement of
$S \to WW$.

Brief comments on the VLQ masses are in order here.
The mass bounds on the VLQs from the direct searches at the Tevatron and the LHC
depend sensitively on the decay channels of the VLQs.
If the main decay mode
includes the third generation quarks,
the bounds
are strong: $M_{\rm VLQ}\gsim 400-600$ GeV~\cite{Okada:2012gy}.
If VLQs mix only with lighter generations, the mass bounds become much less than
$400\gev$ \cite{Okada:2012gy}, which is adopted here.

\begin{figure*}[t!]
    \centering
    \begin{subfigure}[b]{0.5\textwidth}
        \centering
        \includegraphics[width=0.95\textwidth]{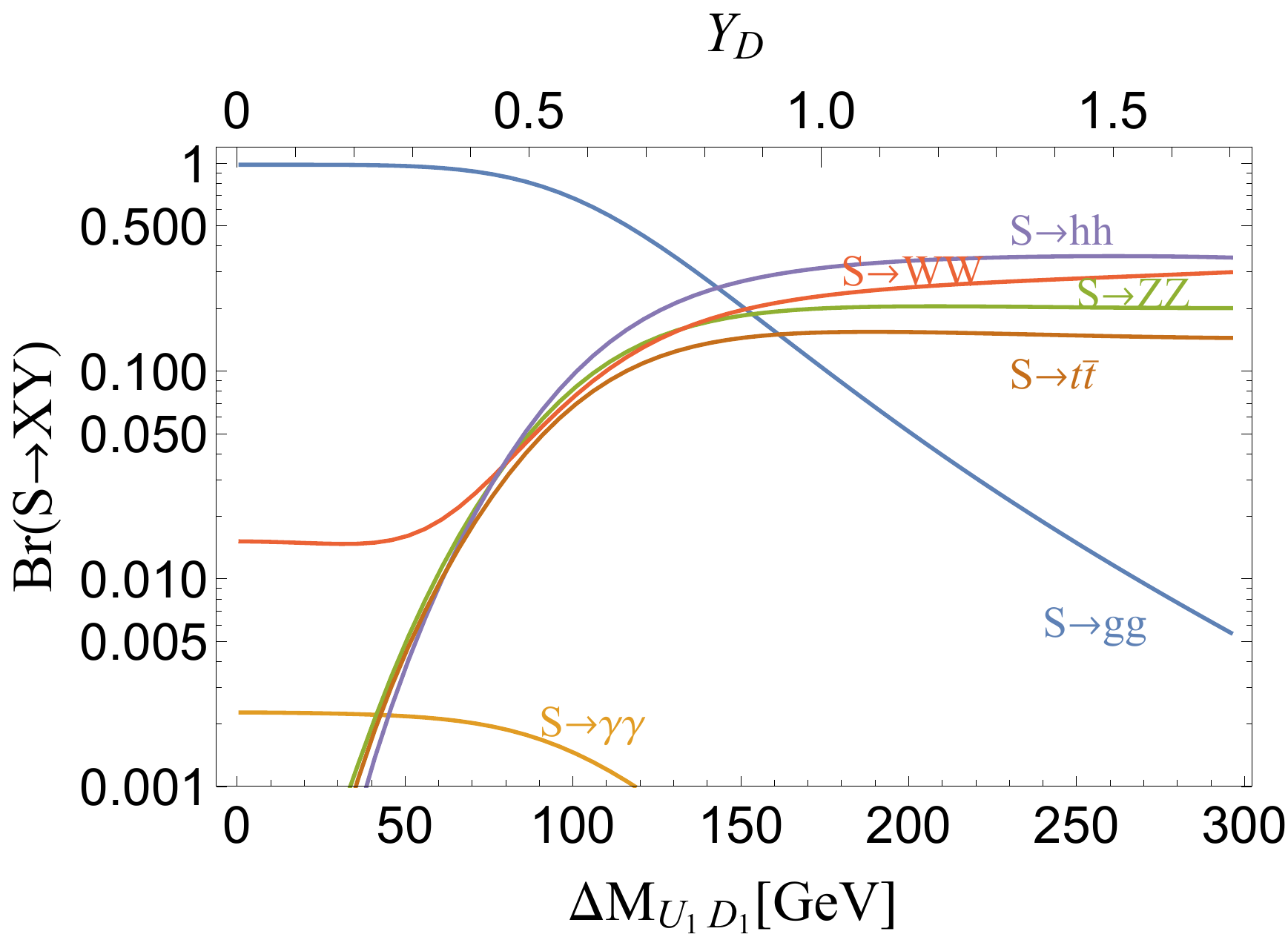}
        \caption{$m_S=500\gev$}
    \end{subfigure}%
    ~
    \begin{subfigure}[b]{0.5\textwidth}
        \centering
        \includegraphics[width=0.95\textwidth]{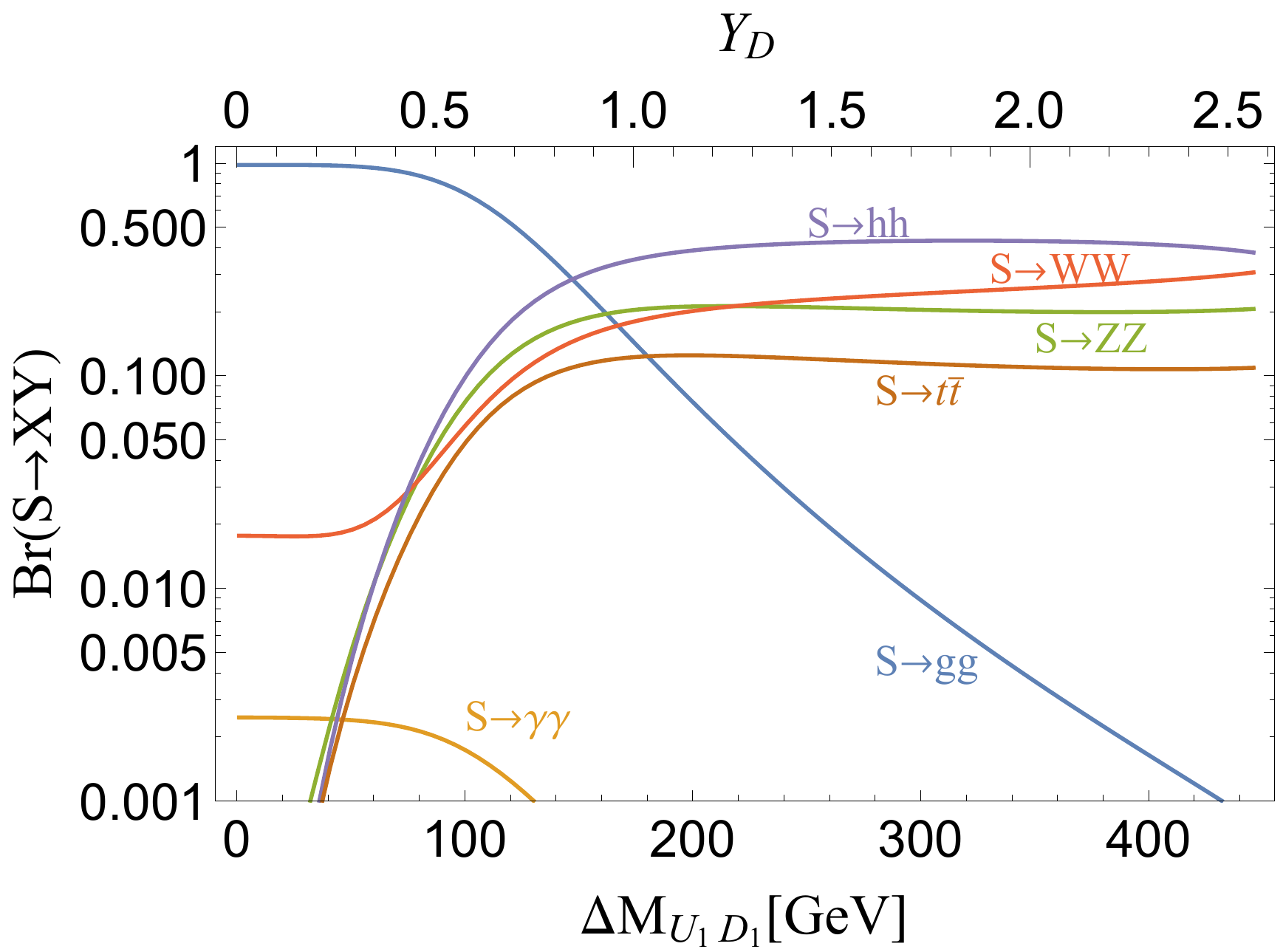}
        \caption{$m_S =750\gev$}
    \end{subfigure}
    \caption{\label{fig-BR}
    Branching ratios of the radiative decays of the singlet scalar $S$
with mass $m_S=500,750\gev$
as functions of $\Dt M_{\mcu_1 \mcd_1}( \equiv M_{\mcu_1} - M_{\mcd_1})$.
For the VLQ masses we set the lightest VLQ mass as $M_{\mcd_1}=0.6 \, m_S$ and assume
$M_\mcq = M_\mcu=M_\mcd$ and $Y_\mcu=0$ 
with varying $Y_\mcd$.}
\end{figure*}

In Fig.~\ref{fig-BR}, we present the branching ratios
of the singlet scalar $S$
as functions of $\Dt M_{\mcu_1 \mcd_1}$, or equivalently of $Y_\mcd$,
along the benchmark parameter line.
We consider two cases, $m_S=500\gev$ and
$m_S = 750\gev$
with $M_{\mcd_1}= 0.6\, m_S$.
When $Y_\mcd =0$ ($Y_\mcu=0$ by setting),
the dominant decay mode is into $gg$ with almost 100\% branching ratio.
The radiative decay of $S$ into $hh$ is certainly prohibited.
In addition there is no radiatively generated $S$-$h$ mixing,
i.e., $s_\eta = 0$,
which forbids the decay of $S \to \ttop$.
The mixing induced decays in $S \to WW,ZZ$
are closed and only the triangle VLQ loop contributions
become relevant.
The next dominant decay mode is into $WW$ with very small branching ratio
of the order of $10^{-3}$.
This is because of the suppression of the longitudinal polarization enhancement
since the $Y_\mcd=0$ condition makes
all of the VLQ masses degenerate
and thus $2 \mca+ \mcb \sim \mathcal{O}(m_V^2/m_S^2)$:
see Eqs.~(\ref{eq:CM:asymptotic}),
(\ref{eq:AM}), and (\ref{eq:BM}).
The reason why $\br(S \to WW)$ is much larger than
$\br(S \to ZZ)$ when $Y_{\mcu,\mcd}=0$ is that the gauge couplings of VLQs 
to the $Z$ boson
are smaller than those to the $W$ boson
with our choice of the electric charges of VLQs.
Note that $\Gm(S \to WW) \gg \Gm(S \to \rr,ZZ)$
is generic in the view of high dimensional operators
in the effective field theory~\cite{Franceschini:2015kwy}.

As $Y_\mcd$ increases,
the decay modes into $hh$, $WW$, $ZZ$ and $\ttop$
all become significant.
For both $m_S=500\gev$ and $m_S=750\gev$ cases,
the $hh$ mode is as important as the $gg$ mode when $Y_\mcd \simeq 0.8$,
and dominant when $Y_\mcd \gsim 0.9$,
followed by the $WW$, $ZZ$, and $\ttop$ modes.
We found that the little hierarchy among $hh$, $WW$, $ZZ$ and $\ttop$ modes
is quite generic with more general parameter setup other than
our benchmark scenario.
In some extreme corners
 of the parameter space such as small $Y_{\mcu,\mcd}$
but large $\Delta_F$,
the $WW$ decay mode is dominant.

\begin{figure}[t]
\begin{center}
\includegraphics[width=.6\textwidth]{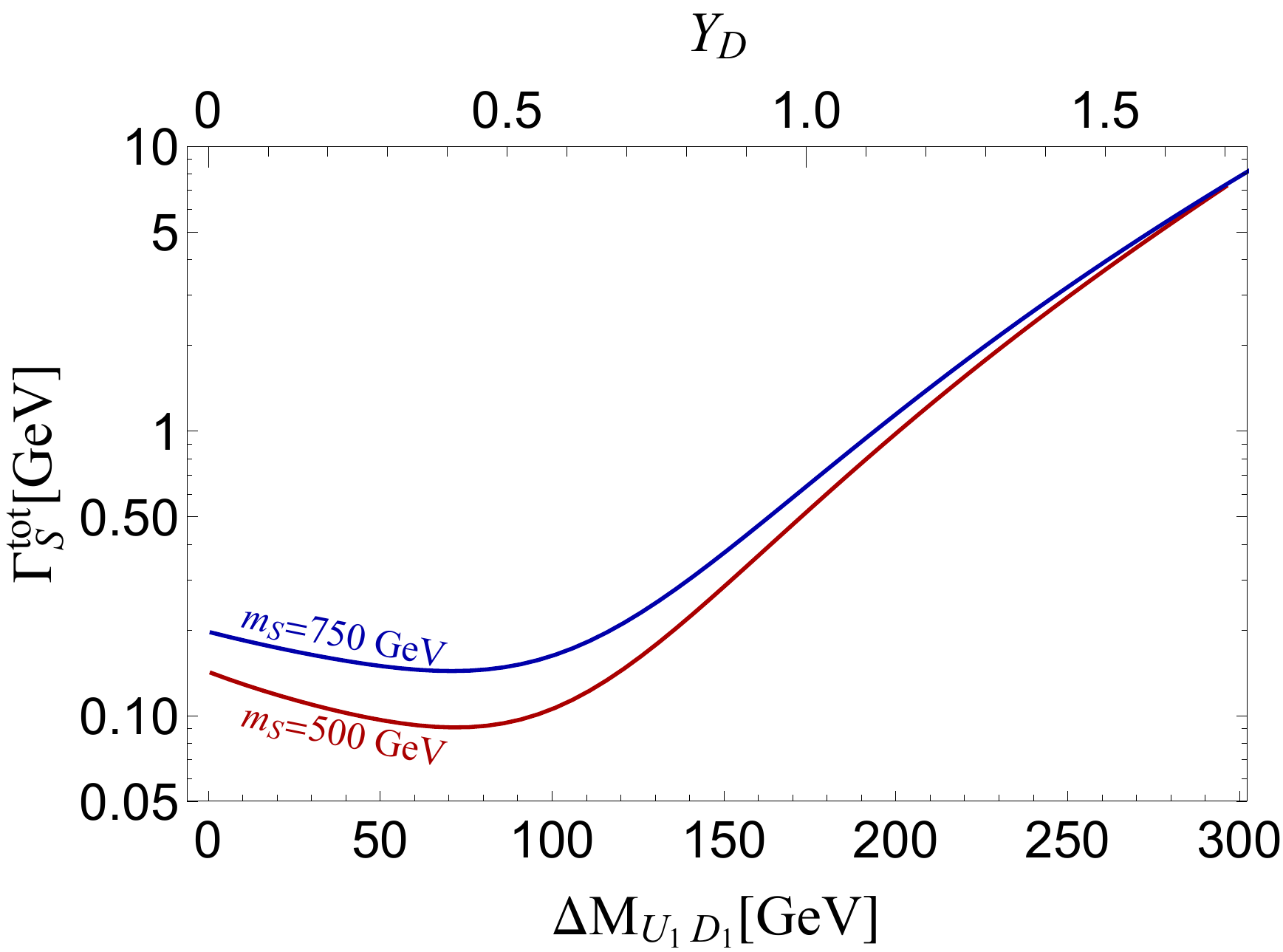}
\end{center}
\caption{\label{fig-gamtot}
\baselineskip 3.0ex
Total decay rate of the singlet scalar $S$
as a function of $\Dt M_{\mcu_1 \mcd_1}$ or $Y_\mcd$
for $m_S=500\gev$ and $m_S=750\gev$.
We take the benchmark parameter line in Eq.~(\ref{eq:benchmark}).
}
\end{figure}

In Fig.~\ref{fig-gamtot},
we show the total decay rate of $S$ as a function of $\Dt  M_{\mcu_1 \mcd_1}$
for $m_S=500\gev$ and $m_S=750\gev$.
When $\Dt  M_{\mcu_1 \mcd_1}=0$,
$\Gm^{\rm tot}_S \sim 0.1 \gev$ for both mass cases.
The singlet scalar is a very narrow resonance.
With increasing $\Dt  M_{\mcu_1 \mcd_1}$,
$\Gm^{\rm tot}_S$ starts decreasing,
which is expected since $\mcu_{1,2}$ and $\mcd_2$
become heavier with
the fixed $M_{\mcd_1}$
and thus make smaller loop corrections.
When $\Dt  M_{\mcu_1 \mcd_1}$ is large enough, however,
$\Gm^{\rm tot}_S$ turns to increase,
reaching about 10 GeV when $\Dt  M_{\mcu_1 \mcd_1}=300\gev$.
The enhancement
compared to the $\Dt  M_{\mcu_1 \mcd_1}=0$ case
is almost by two orders of magnitude.
This is unexpected since the VLQ masses for $\Dt  M_{\mcu_1 \mcd_1}=300\gev$
are much heavier than those for $\Dt  M_{\mcu_1 \mcd_1}=0$.
This shows how dramatical the enhancement of the radiative decays of $S$
can be when there exist sizable mass differences of the VLQs.

\begin{figure*}[t!]
    \centering
    \begin{subfigure}[b]{0.5\textwidth}
        \centering
        \includegraphics[width=0.95\textwidth]{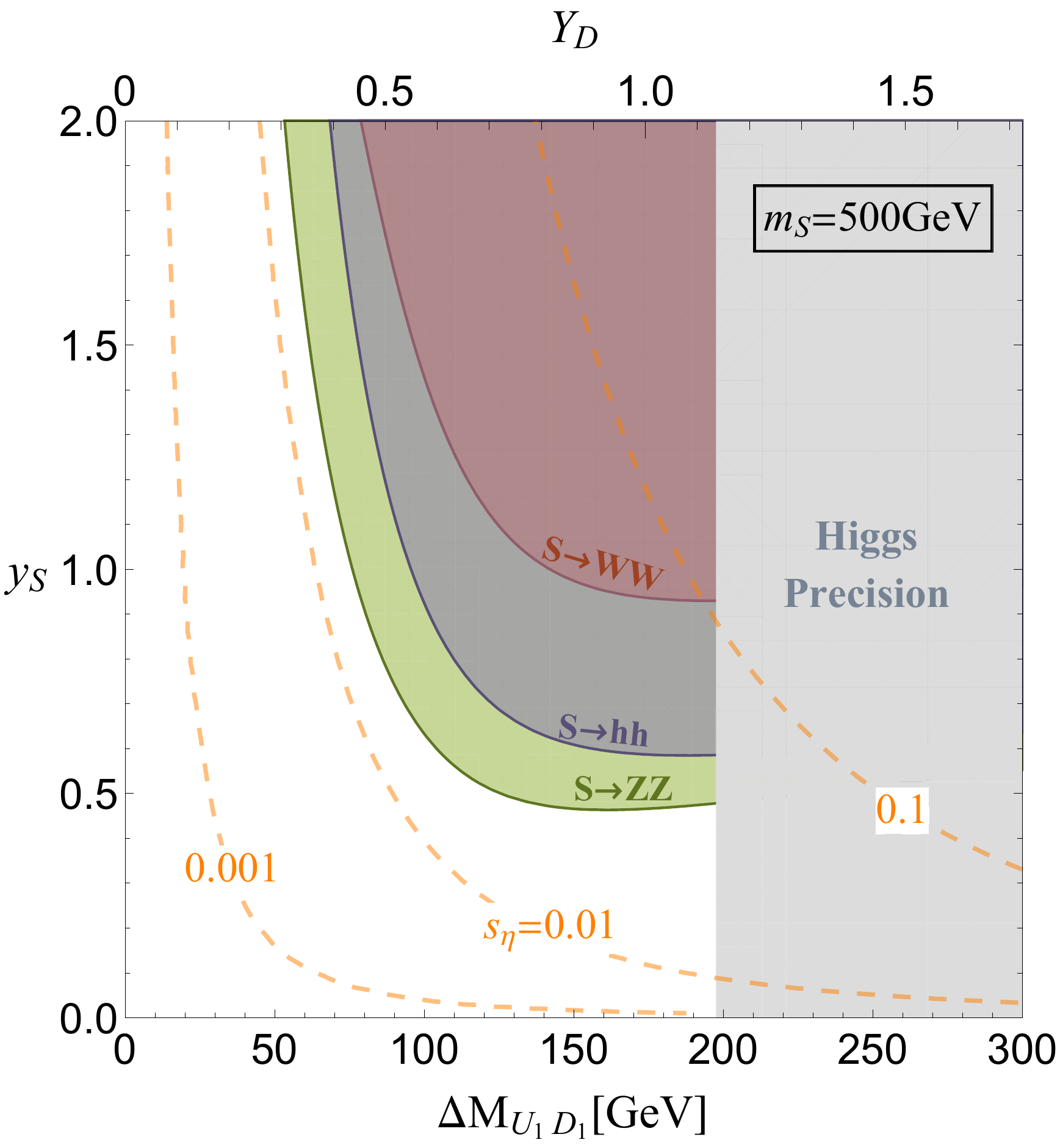}
        \caption{$m_S=500\gev$}
    \end{subfigure}%
    ~
    \begin{subfigure}[b]{0.48\textwidth}
        \centering
        \includegraphics[width=0.95\textwidth]{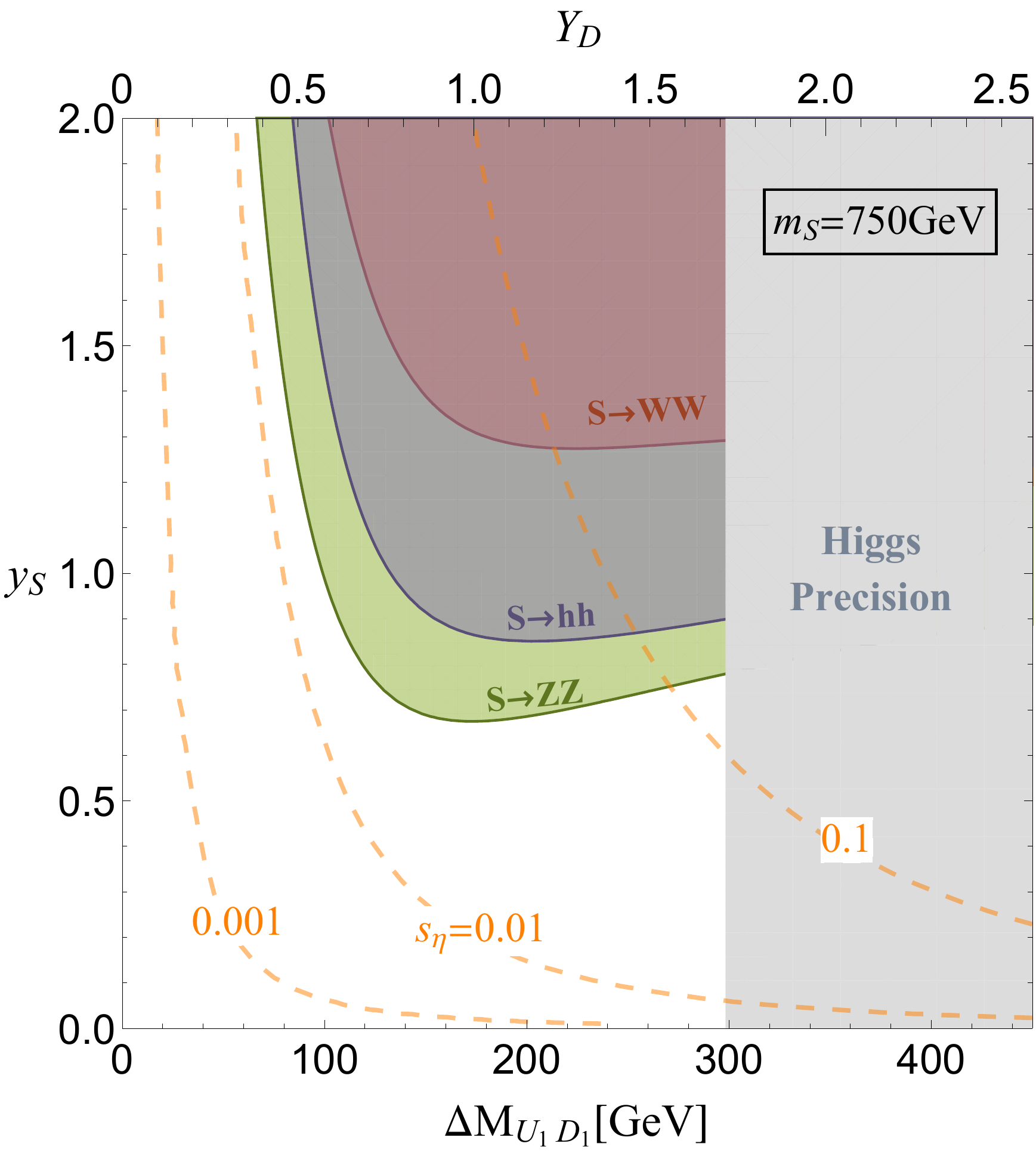}
        \caption{$m_S =750\gev$}
    \end{subfigure}
    \caption{\label{fig-constraints}
The constraints
in the parameter space of $(\Dt M_{\mcu_1\mcd_1}, y_S)$
from the current LHC Higgs data
as well as the $\sqrt{s}=8\tev$ searches for a heavy Higgs
decaying into $WW$,
$ZZ$, and $hh$:
(a) is for $m_S=500\gev$ and (b) is for $m_S=750\gev$.
    }
\end{figure*}

Figure \ref{fig-constraints} presents the 95\% C.L. exclusion region
in the $(\Dt M_{\mcu_1\mcd_1}, y_S)$ parameter plane
by the LHC Higgs precision data as well as
the heavy Higgs search results in the $WW$, $ZZ$, and $hh$ channels.
We also show the contours for $s_\eta$ by dashed (orange) lines.
For the Higgs precision data,
we adopt the global fit results from the ATLAS/CMS combined analysis
for $\kp_V \leq 1$~\cite{kappa:2015}:
$\kp_V = 0.97 \pm 0.060$, $\kp_g = 0.81^{+0.13}_{-0.10}$,
and $\kp_\gm = 0.90^{+0.10}_{-0.09}$.
Note that $\kp_\tau = 0.87^{+0.12}_{-0.11}$
and $\kp_b = 0.57^{+0.16}_{-0.16}$ are consistent within $2\sigma$ but $\kp_t = 1.42^{+0.23}_{-0.22}$
shows some deviation.
For heavy scalar boson searches
with mass $m_S=500$ $(750)\gev$,
the observed 95\% C.L.
upper bounds on $\sg\cdot \br$
at $\sqrt{s}=8\tev$
are
 $200\fb$ ($40\fb$) for $WW$~\cite{Aad:2015agg,Khachatryan:2015cwa},
$43\fb$ ($12\fb$) for $ZZ$~\cite{Aad:2015kna},
and $107.6\fb$ ($34\fb$)
for $hh$~\cite{Aad:2014yja,Aad:2015uka,Aad:2015xja}.
We 
found that the heavy scalar search channels of dijet~\cite{Aad:2014aqa,CMS:2015neg}
and $W\gm/ Z\gm$~\cite{Aad:2014fha} provide weaker
 constraints.
We do not consider the $\ttop$ channel~\cite{Aad:2015fna,Khachatryan:2015sma}
because the current bound ignores the interference with the
continuum background,
which can be very significant~\cite{Jung:2015gta,Carena:2016npr,Barcelo:2010bm}.

The Higgs precision data
exclude large $\Dt M_{\mcu_1 \mcd_1}$,
almost independently of $y_S$:
$\Dt M_{\mcu_1 \mcd_1} \lsim 200 \,(300)\gev$ for $m_S=500\,(750)\gev$ is allowed.
This exclusion mainly comes from the constraint on $\kp_g$
of which the deviation from the SM value is generated from the $S$-$h$ mixing
or the triangle VLQ loops.
When $\Dt M_{\mcu_1 \mcd_1}$ is small,
or equivalently when all of the VLQ masses are almost degenerate,
the opposite signs
between $y_{_{h F_1 F_1}}$ and $y_{_{h F_2 F_2}}$
cause significant cancellation of the $F_1$ and $F_2$ contributions.
Therefore, $\kp_g$ is within the allowed value.
As the VLQ mass difference increases,
the VLQ loop corrections become more important.
The Higgs precision data put an upper bound on $\Dt M_{\mcu_1 \mcd_1}$.
The $\kp_\gm$ is less sensitive since
the dominant contribution to $\kp_\gm$
comes from the $W$ loop.
The $S$-$h$ mixing effect, mainly on $\kp_V$,
is minor since we adopt the Higgs precision data at $2\sg$ level 
such that $s_\eta\lsim 0.5$~\cite{Cheung:2015dta}.

Figure \ref{fig-constraints} shows that
the $ZZ$ channel in the heavy scalar searches puts the strongest bound
for both mass cases.
This is attributed to compatible branching ratios of $WW$, $hh$, and $ZZ$ modes
but 
much smaller LHC upper bounds on $\sg \cdot \br$ for the $ZZ$ mode
because of its clean signal.
The parameter space with large $y_S$ and large $Y_\mcd$
is excluded.
We also present the contours of $s_\eta$
by dashed (orange) lines.
It is clear to see that
the current heavy Higgs searches
put stronger bounds on the $S$-$h$ mixing angle
than the Higgs precision data.
In most parameter space, $s_\eta$ should be less than about 0.01\,(0.05)
for $m_S = 500\,(750)\gev$.
The radiatively generated $S$-$h$ mixing
is significantly constrained by the current LHC data.

\begin{figure}[h] \centering
\begin{center}
\includegraphics[width=.6\textwidth]{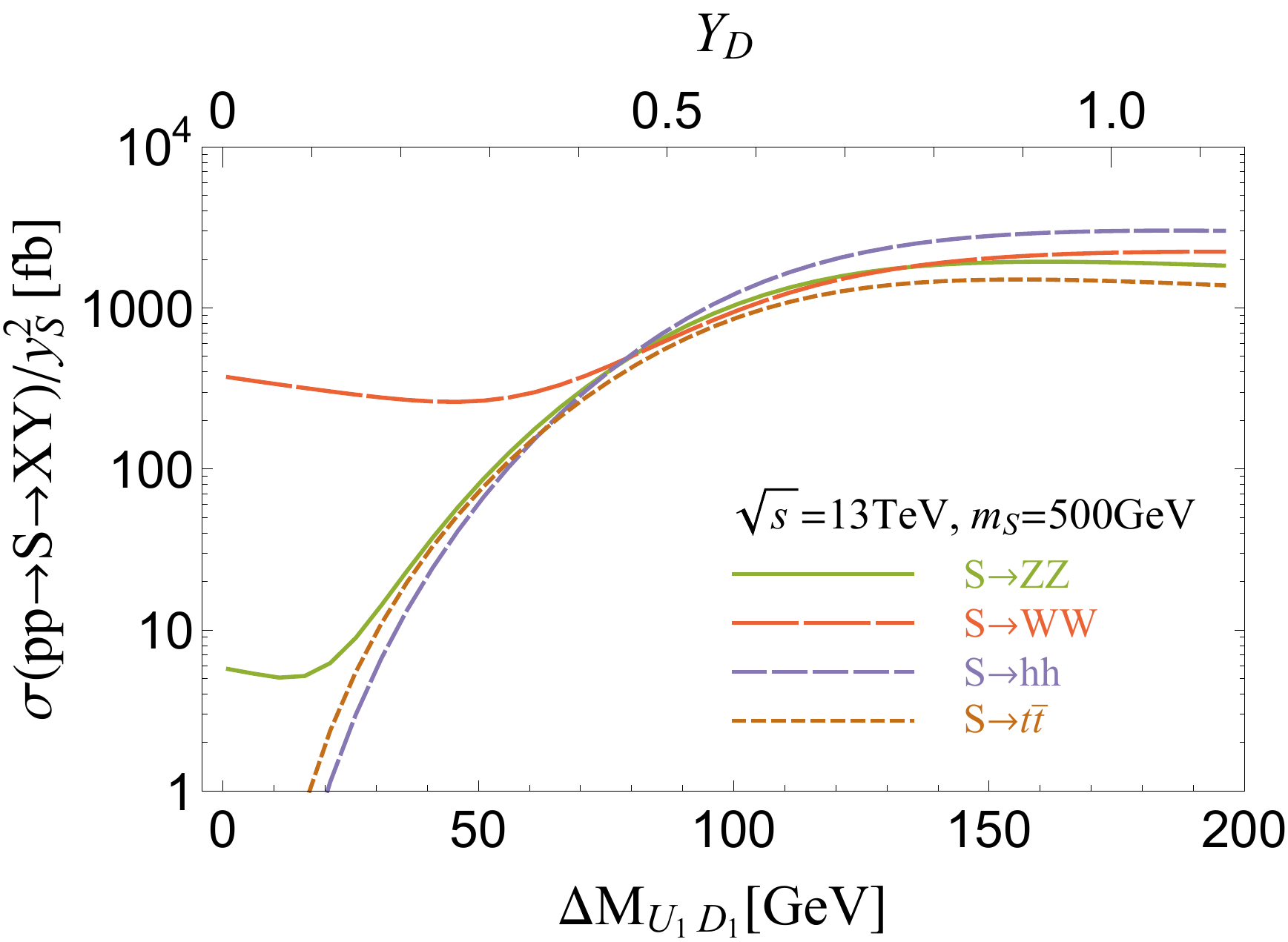}
\end{center}
\caption{\label{fig-500-XS13}
\baselineskip 3.0ex
Cross sections of production and decay of $S$ for the main decay channels 
with $m_S = 500\,{\rm GeV}$ and $\sqrt{s}=13\,{\rm TeV}$ at the LHC. 
The cross sections are normalized by $y_S^2$.
}
\end{figure}

Finally, we show in Fig.~\ref{fig-500-XS13} the cross section times branching ratio 
 $\sigma(pp \to S) \times B(S \to XY)$
as a function of $\Delta M_{\mcu_1\mcd_1}$ with $m_S=500\gev$
at the $13\,{\rm TeV}$ LHC. 
The decay of $S$ into $gg$ is not considered because of the overwhelming QCD background.
We normalize $\sigma \cdot \br$ by $y_S^2$.
Incorporating the current Higgs precision constraint on $\Delta M _{\mcu_1 \mcd_1}$,
we present the results for $\Delta M _{\mcu_1 \mcd_1}$ up to 200 GeV. 
In the whole parameter space,
the $WW$ mode is leading or next-to-leading, 
having $\sg\cdot \br \sim {\cal O}(100-1000)\fb$.
The cleanest search mode, the $ZZ$ one, also has sizable signal rate
about 100 fb
if $\Delta M _{\mcu_1 \mcd_1} \gsim 50\gev$.
The $hh$ channel is also promising with sufficient VLQ mass differences.

\section{Conclusions}
\label{sec:conclusions}

In a simple extension of the SM with an additional singlet scalar field $S$,
we answer the question
whether a unique feature of a heavy scalar boson,
the longitudinal polarization enhancement in its decay into a massive gauge boson pair,
remains at loop level.
In order to focus on the loop induced effects,
we consider a limiting scenario where $S$ does not interact with the SM Higgs boson
at tree level.
Since $S$ decouples from the SM world at tree level,
we introduced vector-like quarks (VLQs) as messengers between $S$ and the SM particles.
In order for the Higgs boson to interact also with the VLQs,
one VLQ doublet and two VLQ singlets are suggested.
There are two up-type VLQs and two down-type VLQs,
$\mcu_{1,2}$ and $\mcd_{1,2}$.
Through the Yukawa couplings of VLQs with $S$ and the Higgs boson,
the VLQs 
generate radiatively the $S$-$h$ mixing as well as
the decays of $S$ into $gg$, $WW$, $ZZ$, and $hh$.

We found that the most required condition
for enhancing the radiative decay rates of $S$
into $WW$, $ZZ$ and $hh$ is the large mass differences of VLQs.
This is contrary to the common expectation
since large mass differences with the fixed lightest VLQ mass
mean heavy VLQs and thus smaller loop corrections.
First the radiatively generated $S$-$h$ mixing is
proportional to the coupling of $h$-$F_i$-$F_i$ ($F=\mcu,\mcd$ and $i=1,2$).
When $M_{F_1} = M_{F_2}$, the opposite signs between 
$h$-$F_1$-$F_1$ and $h$-$F_2$-$F_2$ couplings
cancel the contributions of $F_1$ and $F_2$. 
As $\Dt M_{F_2 F_1}(\equiv M_{F_2}-M_{F_1})$ increases,
the $S$-$h$ mixing angle is enhanced.
The mixing induced decays of $S$ into $ WW$, $ZZ$, $hh$, and $\ttop$ become significant.
Another kind of the VLQ contribution to the radiative decay of $S$ is 
through the triangle VLQ loops.
We showed that the longitudinal polarization enhancement in $S \to WW,ZZ$ 
through the triangular VLQ loops
happens also when the mass differences of the VLQs become large.

In order to illustrate the phenomenological features,
we considered a simple benchmark scenario
where $Y_\mcd$ controls the VLQ mass differences
with the fixed lightest VLQ mass.
Two cases of $m_S=500\gev$ and $m_S=750\gev$ 
are studied.
When $\Dt M_{F F'}=0$, both the $S$-$h$ mixing and the 
longitudinal polarization enhancement in $S\to VV$
vanish, which
makes $S\to gg$ dominant.
The total decay rate is of the order of 0.1 GeV for $m_S\sim 500\gev$.
If $\Dt M_{F F'}$ is sizable such as $Y_\mcd \simeq 0.8$, 
the decay of $S$ into $hh$ becomes as important as that into $gg$.
For $Y_\mcd \gsim 0.8$,
$\br(S \to gg)$ drops rapidly,
and the decays into $hh$, $WW$, $ZZ$, and $\ttop$ become
similarly dominant.
The enhancement of the total decay rate of $S$
is huge, by one order of magnitude when $Y_\mcd =1$.
This is contrary to the naive prediction
that heavier VLQs  running in the loop would cause
smaller loop corrections.

We also presented the 95\% C.L. exclusion regions of $(Y_\mcd, y_S)$
from the current LHC bounds
including the Higgs precision data and the heavy scalar searches 
in the channel of $WW$, $ZZ$, and $hh$. 
Among various Higgs precision data, $\kp_g$
puts the strongest bound on $Y_\mcd$: $Y_\mcd \gsim 1.1$ for $m_S=500\gev$
and $Y_\mcd \gsim 1.7$ for $m_S=750\gev$ are excluded.
The heavy scalar searches also put additional constraints.
In particular the $ZZ$ channel data severely limit the $S$-$h$ mixing angle $\eta$,
more than the Higgs precision data:
$s_\eta \lsim 0.05$  for $m_S=500\gev$
and $s_\eta \lsim 0.1$ for $m_S=750\gev$ are allowed.
In conclusions, 
our loop calculation in a UV model
with a singlet scalar and three VLQ multiplets shows that
the radiative decays of $S$
can be very enhanced when the mass spectrum of the VLQs shows diversity.
Note that the presence of multiple VLQs is crucial for the enhanced radiative decays of $S$
since sizable mass differences among VLQs are required.
Therefore, the persistent searches for a heavy scalar boson at the future LHC
are of great importance in constraining new particles that appear at loop level.

\acknowledgments
K.C. was supported by the MoST of Taiwan under
Grants No. MOST-105-2112-M-007-028-MY3.
S.K. was supported by the National Research Foundation of Korea, 
2015R1D1A1A01058726.
The work of J.S. and Y.W.Y. was supported by the National Research Foundation of Korea, NRF-2016R1D1A1B03932102.

\end{document}